\documentclass[aps,prd,preprint,groupedaddress]{revtex4-1}

\usepackage{graphicx}
\usepackage[table,xcdraw]{xcolor}
\usepackage{xcolor}
\usepackage{amsmath}
\begin{document}

\title{Configuration Entropy for Quarkonium in a Finite Density Plasma}

\author{Nelson R. F. Braga}\email{braga@if.ufrj.br}
\author{Rodrigo da Mata}\email{rsilva@if.ufrj.br }
\affiliation{Instituto de F\'{\i}sica,
Universidade Federal do Rio de Janeiro, Caixa Postal 68528, RJ
21941-972 -- Brazil}


\begin{abstract}  
 In the recent years many examples appeared in the literature where the configuration entropy (CE), introduced by Gleiser and Stamatopoulos,  plays the role of an indicator of stability of  physical systems. It was observed that, comparing states of the same system, the lower is the value of the CE, the more stable is the state.   In this work we investigate the behaviour of  the differential configuration entropy (DCE) , that is appropriate for systems with continuous degrees of freedom, in a new context. We consider quasi-states of quarkonium (a  vector meson made of a heavy quark anti-quark pair) inside a plasma at finite density. It is known that the density increases the dissociation effect  for quasi-particles inside a plasma. So, increasing the density of a thermal medium corresponds to reducing the stability of the quasi-particles. 
In order to investigate how this situation  is translated in the Configutation Entropy context, we use a recently developed  holographic AdS/QCD model for heavy vector mesons.  The quasi-normal modes describing the quasi-states are obtained and the corresponding DCE is calculated. We find, for bottomonium   and  charmonium $1 S$ quasi-states, that the DCE increases with the quark density, or quark chemical potential,  of the medium. This result shows that the DCE works again as an indicator of stability, represented in this case by  the dissociation effect associated with the density. 
\end{abstract}


\maketitle

\section{ Introduction }    

An interesting tool to investigate the stability of physical systems is the configuration 
entropy (CE),   introduced by  Gleiser and Stamatopoulos in refs.\cite{Gleiser:2011di,Gleiser:2012tu} (see also \cite{Gleiser:2013mga}).  An increase in the value of the CE is associated with a decrease in the stability.  Such a behavior was observed in many different physical systems, such as: compact astrophysical objects    
\cite{Gleiser:2015rwa}  and holographic AdS/QCD models \cite{Bernardini:2016hvx,Bernardini:2016qit,Braga:2017fsb,Braga:2018fyc,Bernardini:2018uuy,Ferreira:2019inu}. There are other many interesting applications of configuration entropy in the literature, as for example
 \cite{Alves:2014ksa,Correa:2015vka,Sowinski:2017hdw,Casadio:2016aum,Correa:2016pgr,Braga:2016wzx,Karapetyan:2016fai,Karapetyan:2017edu,Alves:2017ljt,Karapetyan:2018oye,Karapetyan:2018yhm,Gleiser:2018kbq,Colangelo:2018mrt,Lee:2018zmp,Bazeia:2018uyg,Ma:2018wtw,Zhao:2019xle,Ferreira:2019nkz,Bernardini:2019stn,Karapetyan:2019ran} 

 The purpose of the present article is to investigate the application of the configuration entropy to a physical system of great interest currently: heavy mesons inside a quark gluon plasma with finite density.  The insterest in such a system comes from the quark gluon plasma (QGP). 
 This very short-lived state of matter, where quarks and gluons are not confined, is produced in heavy ion collisions and consists of a strongly interacting thermal medium. 
  It is a highly non trivial task  to build up a picture of the QGP  from the particles that reach the detectors.  For reviews about QGP, see for example \cite{Bass:1998vz,Scherer:1999qq,Shuryak:2008eq,CasalderreySolana:2011us}. 
One of the important available sources of information is the abundance of heavy vector mesons, made of $ c {\bar c}$ or  $b {\bar b}$ quarks in the final products of heavy ion collisions. These particles are  partially dissociated in the plasma and their degree of  dissociation depends on the  temperature and density of the plasma. So, it is possible to relate their relative abundance with the  properties of the pre-existing medium.  

It is possible to describe the thermal behavior of heavy vector mesons inside a plasma
using holographic models  \cite{Braga:2015jca,Braga:2016wkm,Braga:2017oqw,Braga:2017bml,Braga:2018zlu}.  The  dissociation of charmonium and bottomnium is represented  in these references as the decrease in the peaks of the thermal spectral functions, that represent the amplitude of finding a particle with a given energy inside the medium.  This dissociation process can alternatively be analyzed through 
the quasinormal modes, that are are normalizable solutions,  with complex frequencies, for the fields that describe the mesons. The real parts of the frequencies are related to the masses and the imaginary parts to the widths of the quasi-states. 
In refs. \cite{Braga:2018hjt,Braga:2019yeh,Braga:2019xwl}, quasi-normal modes for heavy vector mesons were studied using the holographic model of 
 \cite{Braga:2017bml,Braga:2018zlu}.  
 
  The configuration entropy\cite{Gleiser:2013mga}  was motivated by  the  well known  
 information entropy of Shannon \cite{shannon} that is defined, for a discrete variable $x$ that may have the values $x_n$ with  probabilities $p_n$  as
   \begin{equation}
  - \sum_n  \, p_n  \log p_n \,,
 \label{discretepositionentropy}
 \end{equation}
 \noindent and measures the amount of information one gains in getting to know the value of  $x$. Note that the probabilities satisfy the normalization condition: $ \sum_n   p_n = 1$. 
 
 The continous version of eq. (\ref{discretepositionentropy}),  in position space,  reads
  \begin{equation}
S = - \int d^dr \, \epsilon ({\vec r}) \log  \epsilon({\vec r}) \,,
 \label{positionentropy}
 \end{equation}
 where 
 \begin{equation}
  \epsilon ({\vec r}) =\frac{ \vert \rho ({\vec r}) \vert^2}{  \int d^dr \, \vert \rho ({\vec r}) \vert^2}  \,,
  \end{equation}
 is a normalized funtion $\int d^dr  \epsilon({ \vec r }) =1$, called (spatial) modal fraction.  In order to introduce the configuration entropy, one considers the momentum space version, by fourier transforming: 
  \begin{equation}
 {\tilde \rho  } (\vec k )  = \frac{1}{(2\pi)^{d/2}} \int d^dr \,  \rho ({\vec r}) \exp ( -i \vec k \cdot \vec r )   \,. 
 \label{Conjugate}
 \end{equation}
 The  CE is defined as
 \begin{equation}
 \tilde{S} = - \int d^dk \, { \tilde  \epsilon} ({\vec k}) \log  {\tilde  \epsilon} ({\vec k}) \,,
 \label{momentumentropy}
 \end{equation}
 where  
 \begin{equation}
  {\tilde  \epsilon } ({\vec k}) =\frac{  \vert {\tilde \rho } ({\vec k}) \vert^2 }{  \int d^dk \, \vert {\tilde \rho } ({\vec k}) \vert^2}  \,,
  \label{MMF}
  \end{equation}
  is the (momentum space) modal fraction, that is also normalized: $\int d^dk {\tilde  \epsilon } ({\vec k})=1$.
  
Information entropies like $S$ and $\tilde{S}$, based in conjugate variables in the sense of eq. (\ref{Conjugate}), satisfy the so called entropic uncertainty relation \cite{Birula,Braga:2019jqg} that, for this d-dimensional case takes the form:
\begin{equation}
\label{uncertainty}
S + \tilde{S}\geq d(1+log(\pi))\,.
\end{equation}

So, one could guess that a variation of the configuration entropy, defined in momentum space, $\tilde{S}$ could  be associated with a particular variation of the conjugate quantity, $S$, defined in position space. Such a conjecture was investigated recently  in Ref.  \cite{Braga:2019jqg}, where it was found that,  for the case of an anti-de Sitter black hole, when the temperature varies, both $\tilde{S}$   and $S$ vary but their sum remains constant.
 
 For continuum systems, like field theories, it was pointed out in ref. \cite {Gleiser:2018kbq}  that a different kind  of configuration entropy should be used. In order to avoid negavite values, or singularities in the entropy, one should replace the  (momentum)  modal fraction of eq. (\ref{MMF})  by 
 \begin{equation}
  {\widetilde  { D\epsilon } } ({\vec k}) =\frac{  \vert {\tilde \rho } ({\vec k}) \vert^2 }{   \vert {\tilde \rho } ({\vec k}) \vert_{max}^2 }  \,,
  \label{DMF}
  \end{equation}
  where  $ \vert {\tilde \rho } ({\vec k}) \vert_{max}^2 $ is the maximum value of  $ \vert {\tilde \rho } ({\vec k}) \vert^2 $. Then one introduces the differential configuration entropy (DCE) as:
  \begin{equation}
 \tilde{S} = - \int d^dk \, { \widetilde{  D\epsilon}} ({\vec k}) \log  {\widetilde  {D\epsilon}} ({\vec k}) \,,
 \label{Dmomentumentropy}
 \end{equation}
 
 This is the entropy that  we will calculate in this article. For the dual entropy one can follow the same idea and define:
\begin{equation}
  D\epsilon ({\vec r}) =\frac{ \vert \rho ({\vec r}) \vert^2}{  \vert \rho ({\vec r}) \vert_{max}^2} \,,
  \end{equation}
  and a correspondig quantity  
  \begin{equation}
S = - \int d^dr \, D\epsilon ({\vec r}) \log  D\epsilon({\vec r}) \,.
 \label{Dpositionentropy}
 \end{equation}
 We will look at this dual quantity, in order to see if a behaviour similar to the black hole case\cite{Braga:2019jqg}  is found. However it is important to remark that the DCE of eq. (\ref{Dmomentumentropy}) and the dual quantity of eq. (\ref{Dpositionentropy})  are not subject to the relation (\ref {uncertainty}) since $ \vert {\tilde \rho } ({\vec k}) \vert^2 $ and $ \vert \rho ({\vec r}) \vert_{max}^2 $ are not constants.

This article is organized in the following way: in section 2 we review the holographic model for heavy vector mesons in a plasma. In section 3  we  develop the calculation of the differential configuration entropy (DCE)  for charmonium and bottomonium 1S states. Then in section 4 we present and analyse the results obtained and finally in section 5 we present our summary and conclusions. 
  
   \section{Holographic heavy vector mesons at finite density}

Heavy vector mesons are described holographically \cite{Braga:2017bml,Braga:2018zlu} 
by a vector field $V_m = (V_\mu, V_z) (\mu = 0, 1, 2, 3)$,  living in a five dimensional curved space, that is assumed to be dual to the four dimensional  gauge theory current $J_\mu =\bar{\psi} \gamma^\mu\psi$.  The curved five dimensional space  is just an anti-de Sitter space for the case when the mesons are in the vacuum (vanishing temperature and density).  Additionally, there is a scalar background. The action reads
\begin{equation}
I \,=\, \int d^4x dz \, \sqrt{-g} \,\, e^{- \phi (z)  } \, \left\{  - \frac{1}{4 g_5^2} F_{mn} F^{mn}
\,  \right\} \,\,, 
\label{vectorfieldaction}
\end{equation}
where $F_{mn} = \partial_m V_n - \partial_n V_m$. The  background scalar field $\phi(z)$ has the form:  
\begin{equation}
\phi(z)=\kappa^2z^2+Mz+\tanh\left(\frac{1}{Mz}-\frac{\kappa}{ \sqrt{\sigma}}\right)\,.
\label{dilatonModi}
\end{equation}
The parameter  $\sigma$,  with dimension of energy squared,  represents effectively the string tension of the strong quark anti-quark interaction.
 The mass of the heavy quarks is represented by $ \kappa$.  
  The third parameter, $M$,  has a more subtle interpretation.
Heavy vector mesons undergo non hadronic decay processes, when the final state consists of light leptons, like an $e^+ e^-$ pair. In such transitions  there is a  very large mass  change, of the order of the meson mass.  The parameter $ M$  represents effectively the mass scale of such a transition, characterized by a matrix element  $ \langle 0 \vert \, J_\mu (0)  \,  \vert n \rangle = \epsilon_\mu f_n m_n \, $, where $f_n $ is the decay constant, $ \vert n \rangle $ is a meson state  at radial excitation level $n$ with mass $m_n$ , $  \vert 0  \rangle $ is the hadronic vacuum and    $J_\mu$ the hadronic current.  The values that provide the best fit to charmonium and bottomonium spectra of masses and decay constants at zero temperature are respectively
\begin{eqnarray}
\kappa_c&  = 1.2, \sqrt{\sigma_c } = 0.55, M_c=2.2 \, ; \, 
\kappa_b& = 2.45, \sqrt{\sigma_b } = 1.55, M_b=6.2  \,,
  \label{parameters}
  \end{eqnarray}     
\noindent where all quantities are expressed in GeV. 
The geometry dual to a finite temperature medium is in general a black hole one. For the case when the medium additionaly has a finite chemical potential  $\mu$ the black hole has charge\cite{Colangelo:2010pe,Colangelo:2011sr,Colangelo:2012jy}.
In particular, it is a 5-d anti-de Sitter charged black hole space  with metric 
 \begin{equation}
   ds^2 \,\,= \,\, \frac{R^2}{z^2}  \,  \Big(  -  f(z) dt^2 + d\vec{x}\cdot d\vec{x}  + \frac{dz^2}{f(z) }    \Big)   \,,
 \label{metric2}
\end{equation}
where  
\begin{equation}
 f (z) = 1 - \frac{z^4}{z_h^4} - q^2 z_h^2 z^4+q^2 z^6 ,
\end{equation}
 and $f(z_h) = 0$. The relation between the horizon position $z_h$ and the temperature $T$ of the black hole,  is obtained requiring that there is no conical singularity at the horizon:
\begin{equation} 
T =  \frac{\vert  f'(z)\vert_{(z=z_h)}}{4 \pi  } = \frac{1}{\pi z_h}-\frac{  q^2z_h ^5}{2 \pi }\,.
\label{temp}
\end{equation}

The parameter $q$, proportional to the black hole charge, is related to the density of the medium, or  quark chemical  potential, $\mu$ of the gauge theory.  The quantity 
$\mu$ works as the source of correlators of the quark density operator $  \bar{\psi}\gamma^0 \psi \,$. So it should appear in the Lagrangian multiplying the quark density.  
 In the holographic description, the time component $V_0$ of the vector field plays this role. 
So, one considers a particular solution for the vector field   $ V_m $ with only one 
non-vanishing component: $  V_0 = A_0 (z) $  ($V_z =0, V_i = 0$). Assuming that the relation between $q $ and $\mu$ is the same as in the case of no background, that means $ \phi (z)  = 0$, the solution for the time component of the vector field is:
 $ A_{0}(z)=c - qz^2 $, where $c$ is a a constant. Imposing  $A_{0}(0) = \mu $ and $A_{0}(z_h)=0$ one finds: 
\begin{equation}
\mu = q z_h^2  \,.
\label{chemicalpotential}
\end{equation}  
So, specifying both $z_h$ and $q$, the values of the temperature and the chemical potential are fixed and contained into the metric  (\ref{metric2}).

It is interesting to mention that there are many interesting previous studies using holography to describe thermal effects and heavy flavors like for example 
 \cite{Branz:2010ub,Gutsche:2012ez,Fadafan:2011gm,Fadafan:2012qy,Fadafan:2013coa,Afonin:2013npa,Mamani:2013ssa,Hashimoto:2014jua,Liu:2016iqo,Liu:2016urz,Ballon-Bayona:2017bwk,Dudal:2018rki,Gutsche:2019blp,Bohra:2019ebj,Zhang:2019qhm,MartinContreras:2019kah}.

 \section{Configuration entropy of the  heavy mesons} 
 \subsection{Energy Density} 
  
 The quantity that is relevant for the determination of the configuration entropy of  heavy mesons is the energy density, that is the $ T_{00} $ component of the energy momentum tensor. We assume that in this 
 phenomenological model $ T_{mn} $ is obtained from the action in the same way as  in general relativity.  That means,  writing the action as $   \int d^4x dz \, \sqrt{-g} {\mathcal L} $ the energy momentum tensor has the form:
\begin{equation}\label{EnTe}
T_{mn}(z)=\frac{2}{\sqrt{-g}}\!\left[ \frac{\partial(\sqrt{-g}\mathcal{L})}{\partial g^{mn}}-\frac{\partial}{\partial x^{p}}\frac{\partial(\sqrt{-g}\mathcal{L})}{\partial\left(\frac{\partial g^{mn}}{\partial x^{p}} \right)} \right] \,. 
\end{equation}
 So, for the action (\ref{vectorfieldaction}) the energy density for the vector field is 
\begin{equation}
\!\!\!\!\!\!\rho (z)\!=\!\frac{e^{- \phi (z)  } }{g^{2}_{5}}\!\left[g_{00}\!\left(\frac{1}{4}g^{mp}g^{nq}F_{mn}F_{pq}\right)\!-\!g^{mn}F_{0n}F_{0m}\right]\,.
\end{equation}  
Considering the metric (\ref{metric2})  and  a solution corresponding to a meson at rest  in the   $x^\mu$ directions $V_{\mu}=\eta_{\mu}v(p,z)e^{-i\omega t}$, with $\eta_{\mu}=(0,1,0,0)$, the energy density takes the form 
\begin{equation}\label{rho}
 \!\!\!\!\!\!\rho (z)\!=\!\frac{z^2 e^{- \phi (z)  } }{2 R^2 g^{2}_{5}}\!\left[|\omega|^2 |v|^2 + f^2|\partial_z v|^2\right]\,.
\end{equation}
In order to obtain the energy density for a meson inside the plasma, one has to find the solution for the field $v$ representing the corresponding quasistate and plug it into  eq. (\ref{rho}). 
At zero temperaure, states are represented holographically by normalizable solutions of the gravity field equations. This type of solutions are called normal modes and satisfy trivial boundary conditions. On the other hand, at finite temperature,  the solutions that represent the quasistates are the so called quasinormal modes, that are also normalizable solutions of the field $ v  $ but satisfy  non trivial boundary conditions. 
At finite $T$ there is an event horizon at $z = z_h $  where one has to impose infalling boundary conditions. Additionally,  the normalizability condition requires that the fields vanish at the boundary $ z = 0$.  Satisfiying both conditions requires, in general, solutions corresponding to complex frequencies $\omega $. The real part, $\operatorname{Re}(\omega )$, is related to the thermal mass and the imaginary part, $\operatorname{Im}(\omega )$, is related to the thermal width. 
We will see in the next section  how to obtain these solutions.

\subsection{ Quasinormal modes }

As in the previous section,  we consider $V_z = 0$ and  $V_{\mu}=\eta_{\mu}v(p,z)e^{-i\omega t}$, with $\eta_{\mu}=(0,1,0,0)$. Introducing the electric field component $E = \omega V_1$, the equations of motion coming from action (\ref{vectorfieldaction}) with the metric(\ref{metric2}) take the form:
\begin{equation}
\label{eqzz}
E''+\left(\frac{f'}{f}-\frac{1}{z}-\phi'\right)E'+\frac{\omega^{2}}{f^2} E =0\,, 
\end{equation} 
where (') represents derivative with respect to the radial z coordinate.

One has to impose the normalizability condition at $z=0$ and the infalling condition at $z=z_h$. 
It is convenient, in order to impose the boundary conditions at the horizon,  to re-write the field equations in such a way that they separate into a combination of  infalling  and outgoing waves. One introduces the coordinate $r_{*}$, implicitly defined by the relation $\partial_{r_{*}} = - f(z)\partial_{z}$ with $r_{*}(0)=0$,  for z in the $0 \leq z \leq zh$. In addition let us introduce the field 
\begin{equation}
\psi = e^{-\frac{B (z)}{2}}E \,,
\label{anzats}
\end{equation}
with $ B(z) = \log(z/R) + \phi$.  Then, Eq. (\ref{eqzz}) reduce to the form:
\begin{equation}
\label{Sequation}
\partial^{2}_{r_*}\psi +\omega^{2}\psi = U\psi \,.
\end{equation}
The potential $U(z)$, obtained this way, diverge at $z=0$ so one  must impose $\psi(z=0)=0 $. At the horizon $U(z=z_h)=0$ so one expects to find  \textit{infalling} $\psi = e^{-i\omega r_*}$ and \textit{outgoing} $\psi = e^{+i\omega r_*}$ wave solutions for equation (\ref{Sequation}). Only the first kind of solutions are physically allowed. The Sch\"{o}dinger like  equation can be expanded near the horizon leading to the following expansion the field solution:
\begin{equation}
\psi=e^{-i\omega r_*(z)} \left[1+a^{(1)}\left(z-z_h\right)+a^{(2)}\left(z-z_h\right)^2+\dots\right] \,.
\label{expansion}
\end{equation}
One can solve recursively for $a^{(n)}$. The first coefficient obtained is:
\begin{equation}
a^{(1)}=   \frac{\left(2 - q^2 z_h^6 \right) \left(z_h \left(\frac{k^2}{2-q^2
   z_h^6}+2 \kappa ^2\right)-\frac{\text{sech}^2\left(\frac{\kappa
   }{\sqrt{\sigma }}-\frac{1}{M z_h}\right)}{M
   z_h^2}+\frac{1}{z_h}+M\right)}{2 \left(q^2 z_h^6+i \omega 
   z_h-2\right)} \,.
\end{equation}

This expansion leads to the following form for the  infalling boundary conditions for the field and it's derivative at the horizon: 
\begin{eqnarray}
E(z_h) &=&  e^{-i\omega r_*(z_h) + \frac{B(z_h)}{2}} \,, \\
E^{'}(z_h)  &=&  \left(-i\omega r'_{*}(z_h) + \frac{B'(z_h)}{2}+a^{(1)}_j\right) E(z_h) \,.
\end{eqnarray} 
Then one solves eq. (\ref{eqzz}) numerically integrating from the horizon, using  a method that consists of imposing these infalling boundary conditions and search for complex frequencies that provide solutions vanishing on the boundary: $E(z=0)=0$. The results are the  quasinormal frequencies and the corresponding solutions  are the quasinormal modes, that represent the heavy meson quasi-states in the thermal medium. 
  
\subsection{ Entropy }
   
The solutions for the gravity fields that holographycaly describe the heavy vector mesons are complex. So, the actual form of the Lagrangian density is $F^{*}_{mn} F^{ mn}$.   
The Configuration entropy is calculated form the solutions  $v(p,z)$ corresponding to the  the quasinormal modes $v_n (z)$,  described in the previous section. One considers the 
  Fourier transform of the energy density $ \rho (z)$ in coordinate $z$:  ${\tilde \rho (k)}  $. It is convenient,  for the computation of the CE, to split ${\tilde \rho} (k) = \left( C(k) -  iS(k)\right) / \sqrt{2\pi} $, where
\begin{eqnarray}
C(k)&=&\int_{0}^{z_h}\rho(z)\cos({kz})dz \,,\\
S(k)&=&\int_{0}^{z_h}\rho(z)\sin({kz})dz \,.
\end{eqnarray}
In terms of these components,  the  modal fraction reads:
\begin{equation}
{\widetilde {D\epsilon}}(k)=\frac{S^2(k)+C^2(k)}{ \left[ S^2(k)+C^2(k) \right]_{max}}\,.
\end{equation}
For this one dimensional case, the  CE (\ref{momentumentropy}) reads 
\begin{equation}\label{CE}
{\tilde S}  = -\int^{\infty}_{-\infty}  \widetilde {D\epsilon}(k)\log{ \left[ \widetilde {D\epsilon}(k)\right]  } \,dk\,.
\end{equation}

\begin{figure}[t]\label{BBresults}
\centering
\includegraphics[scale=0.5]{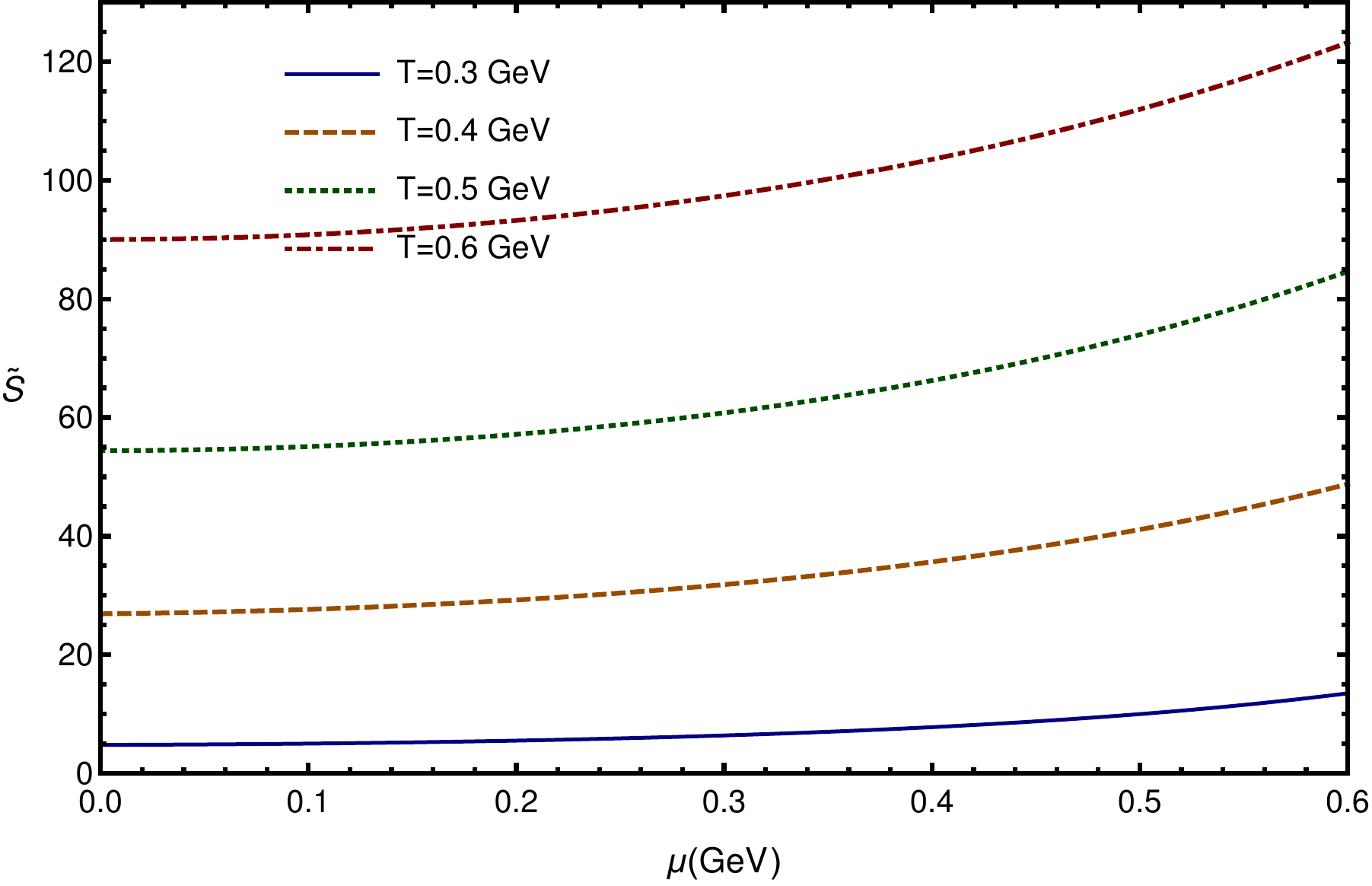}
\caption{Differential configuration entropy ${\tilde S}$ as a function of the chemical potential for bottomonium   $\Upsilon$   at  temperatures:  $T$   = 300 (blue, solid line) , 400 (brown, slashed line) ,  500 (green, dotted line) and  600 MeV (red, dash-dotted line).}
\end{figure}

\begin{figure}[t]\label{CCresults}
\centering
\includegraphics[scale=0.63]{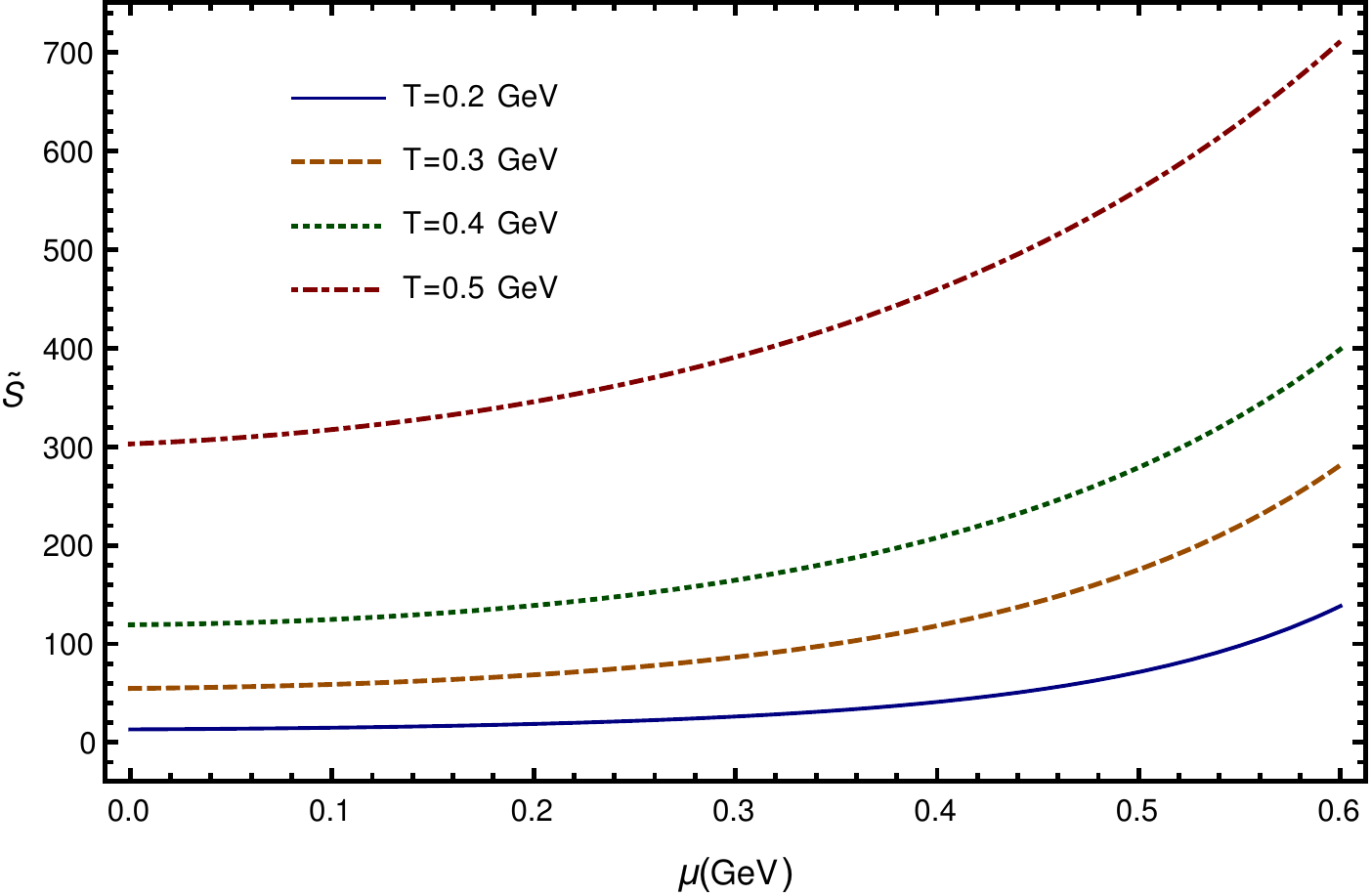}
\caption{Differential configuration entropy ${\tilde S}$ as a function of the chemical potential for charmonium  ${\tilde S}$  at  temperatures:  $T$   = 200 (blue, solid line) , 300 (brown, slashed line) ,  400 (green, dotted line) and  500 MeV (red, dash-dotted line).}
\end{figure}

\begin{figure}[h!]\label{dualBotomonium}
\centering
\includegraphics[scale=0.48]{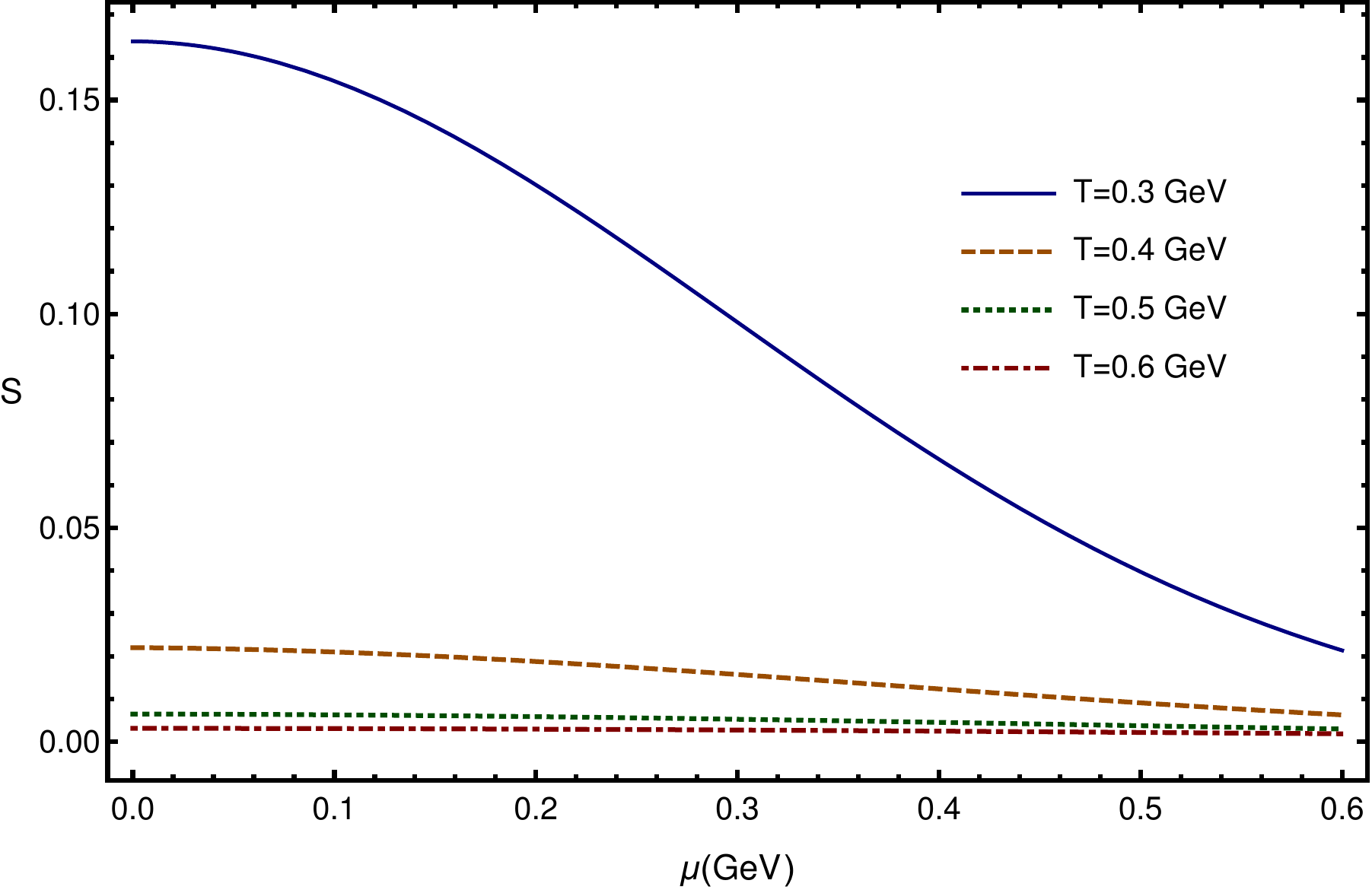}
\caption{  Dual  entropy $ S$ as a function of the chemical potential for bottomonium   $\Upsilon$   at  temperatures:  $T$   = 300 (blue solid line) , 400 (brown slashed line) ,  500 (green dotted line) and  600 MeV (red dash-dotted line). }
\end{figure}

\begin{figure}[h!]\label{dualCharmonium}
\centering
\includegraphics[scale=0.63]{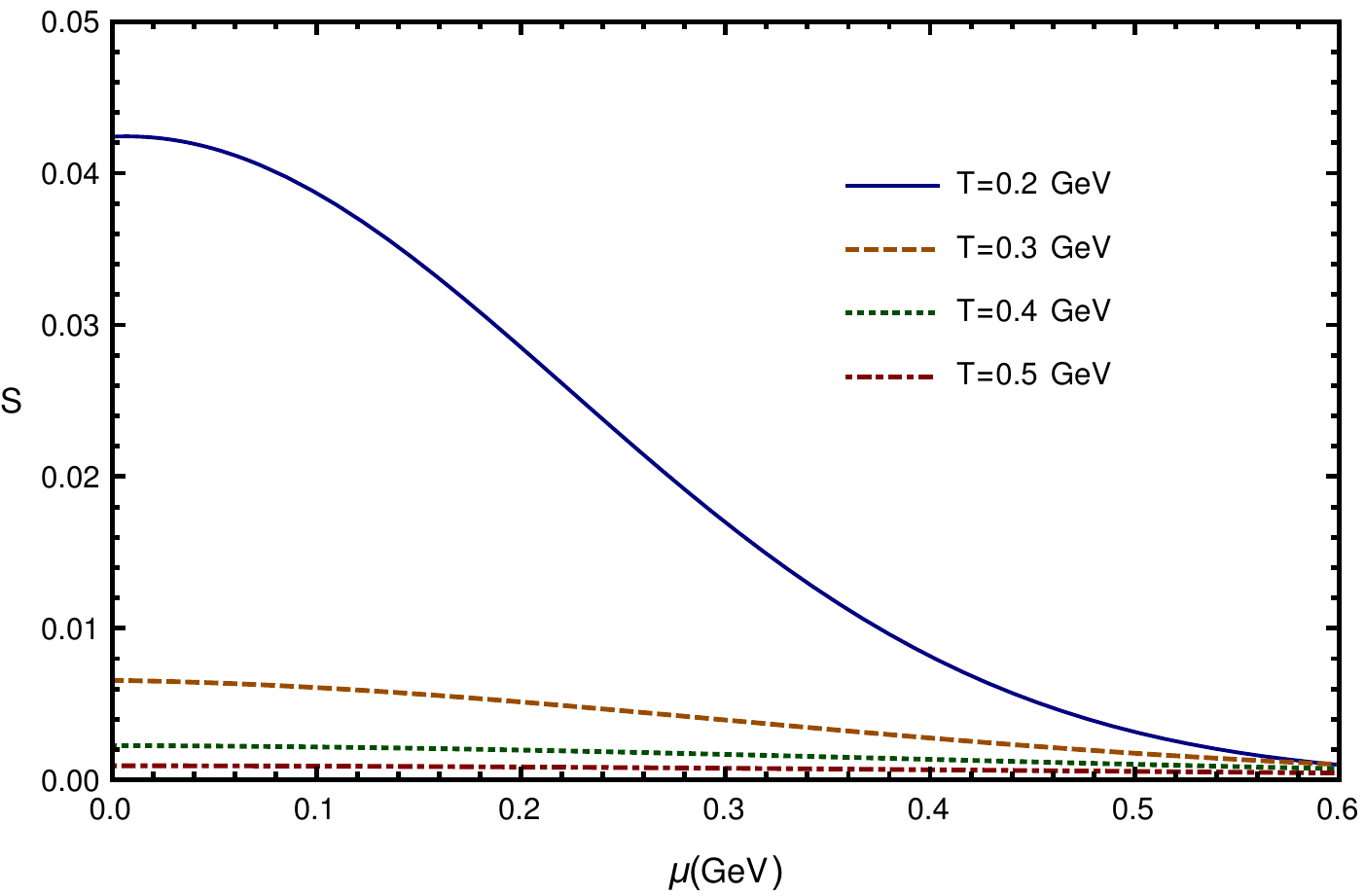}
\caption{  Dual entropy $ S$ as a function of the chemical potential for charmonium  ${\tilde S}$  at  temperatures:  $T$   = 200 (blue solid line) , 300 (brown slashed line) ,  400 (green dotted line) and  500 MeV (red dash-dotted line).  }
\end{figure}

\section{Results} 
We calculated the DCE  for bottomonium and charmonium 1S states, considering four representative temperatures. The dissociation occurs in different temperatures for these two heavy mesons. Charmonium  1S  state, the  J/$\psi$ ,  dissociates at lower values of T, so we chose temperatures  ranging from 200 Mev to 500 MeV, while for bottomonium  1S state,   the $\Upsilon$,  that dissociates at higher temperatures, we chose a range from 300 MeV to 600 MeV.  In figure {\bf 1 } we show plots for the  differential configuratio entropy DCE  for the bottomonium 1S state. Then in figure {\bf 2} we present  similar plots for the charmonium 1S state. In both cases one clearly sees  that the DCE increases with the temperature and the density. The variation with the temperature was analysed before in ref. 
\cite{Braga:2018fyc}, where the same bahavior was found. It is known that the dissociation effect of vector mesons inside the plalma is enhanced by the tempearture and by the density. As the  
 the density and (or) the temperature of the plasma  increase, the dissociation degree of  heavy mesons increases. So, they become more unstable in the sense of their tendency to ``melt" in the plasma.  So, the results shown in figures {\bf 1} and {\bf 2} for the DCE are consistent with the interpretation that instability  corresponds  to an increase in the value of this quantity.

 We also calculated the dual quantity  $ S (\mu ) $ from eq. (\ref{Dpositionentropy}), that is the position space conjugate of the DCE, we found a result that is similar to the one obtained in  \cite{Braga:2019jqg}. That means  $ S (\mu ) $  has the opposite behavior, decreasing with $\mu$.  
 We show in  figures {\bf 3}  and {\bf 4}  the value of  $ S (\mu ) $ for bottomonium and charmonium,  respectively, for the same temperature ranges. 
 
\begin{figure}[h!]\label{scalingBot}
\centering
\includegraphics[scale=0.47]{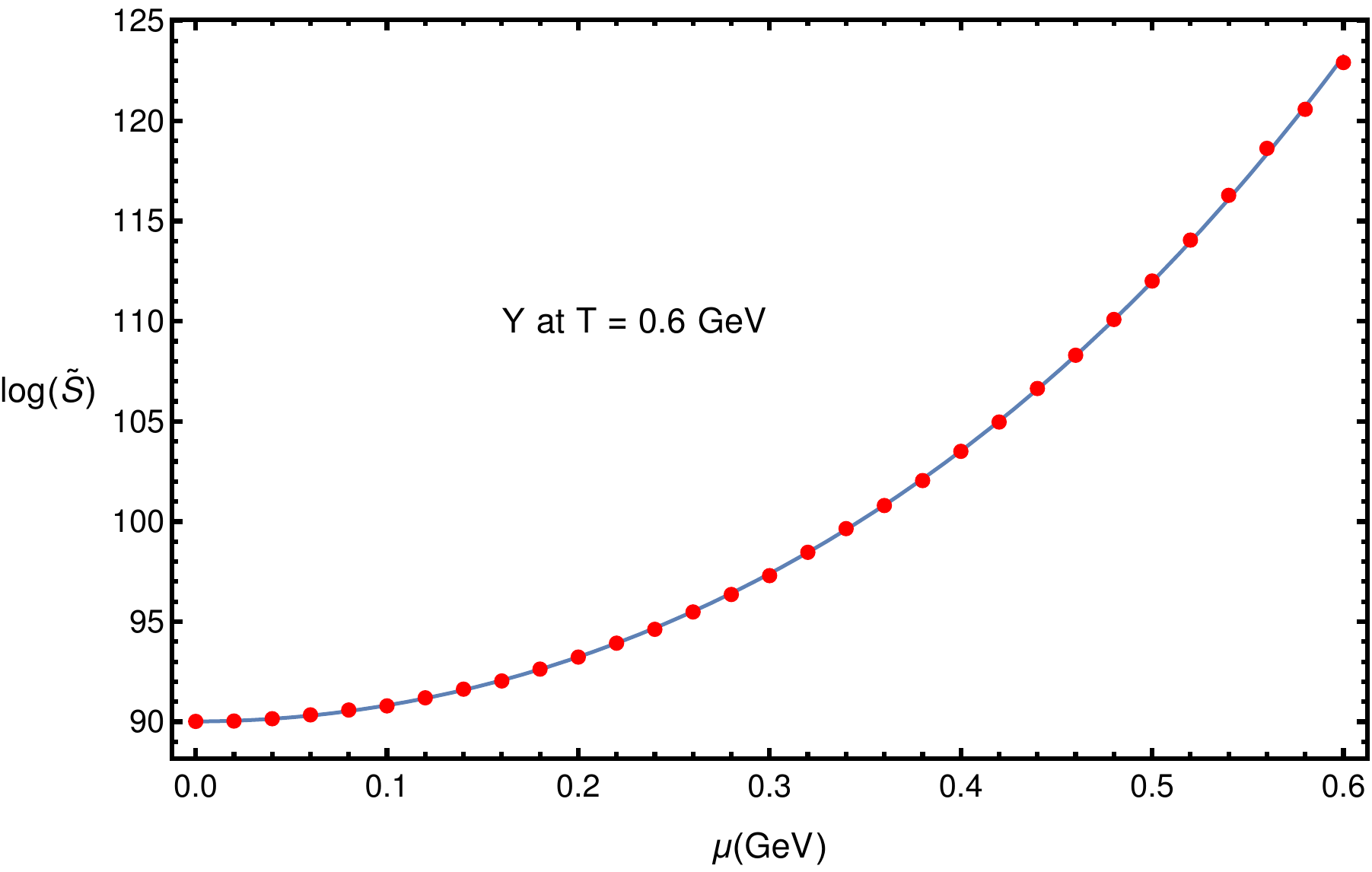}
\caption{  Logarithm of the differential configuration entropy as a function of  $\mu$ for   $\Upsilon$ at temperature 600 MeV. The dots are the values for log(${\tilde S}$)  obtained form the model and the continuous line is the second order polynomial adjust.   }
\end{figure}

\begin{figure}[h!]\label{scalingCharmonium}
\centering
\includegraphics[scale=0.60]{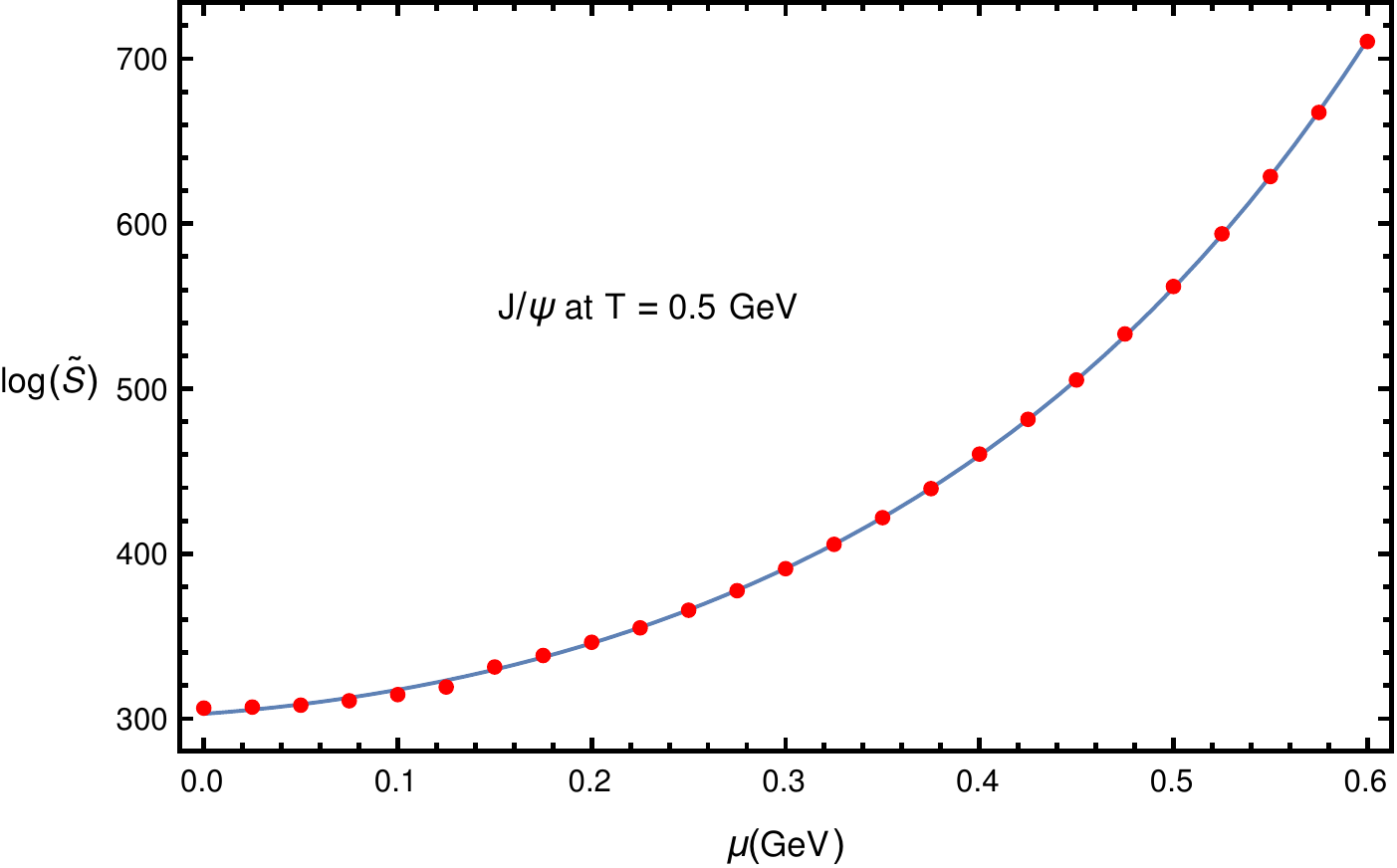}
\caption{  Logarithm of the differential configuration entropy as a function of  $\mu$ for   J/$\psi$ at temperature 500 MeV. The dots are the values for log(${\tilde S}$)  obtained form the model and the continuous line is the second order polynomial adjust.    }
\end{figure}

 It is interesting to investigate if  the dependence of the DCE on the density $\mu$ can be expressed in the form of a  scaling law. As an illustration, we plot in figures  {\bf 5}  and  {\bf 6} the logarithm of the DCE as a function of the density for  $\Upsilon$ at T =  600 MeV and 
  J/$\psi$ at T =  500 MeV, respectively. 
 From the analysis of this kind of plot, for different temperatures, one finds  that there is an  approximate   scaling law  of the  form
\begin{equation}\label{scalinglaw}
log({\tilde S}) = c_0  + c_1 \mu + c_2 \mu^{2},
\end{equation}
\noindent where the coefficients $c_0, c_1 , c_2$ depend on the temperature. We show on tables { \bf 1}  and  {\bf 2}   the values obtained for these parameters at different temperatures and the $R^{2}_{Adj}$ for the   polynomial approximation for   J/$\psi$  and  $\Upsilon$, respectively.

\begin{table}[h]\label{data1}
\centering
\begin{tabular}{|l||l||l||l||l|}
\hline
\multicolumn{5}{|c|}{Scaling coefficients for J/$\psi$ differential configurational entropy adjust}\\ \hline \hline
T (GeV)&    $ c_0$&$ c_1 (GeV)^{-1}$ & $  c_2(GeV)^{-2}$ & $R^{2}_{Adj} $\\ \hline
0.2 & 2.586 $\pm$0.022   & 0.66 $\pm$0.11      & 5.41 $\pm$0.13     &  0.999923 \\ \hline
0.3 & 4.0044 $\pm$0.0042 & 0.319 $\pm$0.024    & 4.011 $\pm$0.033   & 0.999986  \\ \hline
0.4 & 4.7831 $\pm$0.0015 & 0.1284 $\pm$0.0097  & 3.136 $\pm$0.014   & 0.999996 \\ \hline
0.5 & 5.7135 $\pm$0.0026 & 0.279 $\pm$0.017    & 1.906 $\pm$0.025   &  0.9999988 \\ \hline
\end{tabular}
\caption {Coefficients $ c_0$, $ c_1$ and $ c_2$ of  eq. \ref{scalinglaw} for   for J/$\psi$ meson at different temperatures  .}
\end{table}

\begin{table}[h]\label{data2}
\centering
\begin{tabular}{|l||l||l||l||l|}
\hline
\multicolumn{5}{|c|}{Scaling coefficients for $\Upsilon$ differential configurational entropy adjust}\\ \hline \hline
T (GeV)&   \centering $ c_0 $&$  c_1 (GeV)^{-1}$ & $  c_2(GeV)^{-2}$ &$ R^{2}_{Adj} $\\ \hline
0.3 & 1.5697 $\pm$0.0015   & 0.184 $\pm$0.0010    & 2.562 $\pm$0.015   &  0.999995 \\ \hline
0.4 & 3.2928 $\pm$0.0030   & 0.130 $\pm$0.021     & 1.437 $\pm$0.031    & 0.999975  \\ \hline
0.5 & 3.9967 $\pm$0.0022   & 0.002 $\pm$0.015     & 0.006 $\pm$0.024    & 0.999986  \\ \hline
0.6 & 4.50002 $\pm$0.00055 & 0.00018 $\pm$0.0040  & 0.8691 $\pm$0.0062  &  0.999999 \\ \hline
\end{tabular}
\caption {Coefficients $ c_0 $, $  c_1$ and $  c_2$ of eq. \ref{scalinglaw}  for  $\Upsilon$ meson at different temperatures.}
\end{table}

\section{Conclusions}
We studied in this article  the variation of the differential configuration entropy DCE for the 1S states of charmonium (the $ J/\psi$) and bottomonium (the $\Upsilon$)  as a function of the quark density of the medium, for different temperatures. The results obtained show that the entropy increases with the temperature and with the density. This behaviour is consistent with the expectation that an increase in the instability of a physical system should correspond to an increase in the DCE. The higher are the values of temperature and (or)  density  the higher is the probability that the heavy vector meson will dissociate in the thermal medium. So, increasing the temperature or the the density, the quasistates become more unstable against  ``melting``  in the plasma. This analysis provides one more example of  how the DCE  carachterizes the stability of physical systems.

\noindent {\bf Acknowledgments:}  The authors are supported by  CNPq - Conselho Nacional de Desenvolvimento Cientifico e Tecnologico (N.B by grant  307641/2015-5 and R. M. by a graduate fellowship). This work received also support from  Coordenação de Aperfeiçoamento de Pessoal de Nível Superior - Brasil (CAPES) - Finance Code 001.


\begin{thebibliography}{ABC}



\bibitem{Gleiser:2011di} 
  M.~Gleiser and N.~Stamatopoulos,
  Phys.\ Lett.\ B {\bf 713}, 304 (2012)
  [arXiv:1111.5597 [hep-th]].
  
\bibitem{Gleiser:2012tu} 
  M.~Gleiser and N.~Stamatopoulos,
  Phys.\ Rev.\ D {\bf 86}, 045004 (2012)
  [arXiv:1205.3061 [hep-th]].


\bibitem{Gleiser:2013mga} 
  M.~Gleiser and D.~Sowinski,
  Phys.\ Lett.\ B {\bf 727}, 272 (2013)
  [arXiv:1307.0530 [hep-th]].


\bibitem{Gleiser:2015rwa} 
  M.~Gleiser and N.~Jiang,
  Phys.\ Rev.\ D {\bf 92}, no. 4, 044046 (2015)
  [arXiv:1506.05722 [gr-qc]].

 
\bibitem{Bernardini:2016hvx} 
  A.~E.~Bernardini and R.~da Rocha,
  Phys. Lett. B {\bf 762}, 107 (2016) 
  [arXiv:1605.00294 [hep-th]].
    
    
    
    
\bibitem{Bernardini:2016qit}
  A.~E.~Bernardini, N.~R.~F.~Braga and R.~da Rocha,
  Phys. Lett. B {\bf 765}, 81  (2017), 
  [arXiv:1609.01258 [hep-th]].
 
 \bibitem{Braga:2017fsb} 
  N.~R.~F.~Braga and R.~da Rocha,
  Phys.\ Lett.\ B {\bf 776}, 78 (2018)
  [arXiv:1710.07383 [hep-th]].  
 
\bibitem{Braga:2018fyc} 
  N.~R.~F.~Braga, L.~F.~Ferreira and R.~Da Rocha,
  Phys.\ Lett.\ B {\bf 787}, 16 (2018)
  [arXiv:1808.10499 [hep-ph]].
  
 \bibitem{Bernardini:2018uuy}
  A.~E.~Bernardini and R.~da Rocha,
  Phys.\ Rev.\ D {\bf 98} (2018) 126011
  [arXiv:1809.10055 [hep-th]].
 
     
\bibitem{Ferreira:2019inu} 
  L.~F.~Ferreira and R.~Da Rocha,
  Phys.\ Rev.\ D {\bf 99}, no. 8, 086001 (2019)
  [arXiv:1902.04534 [hep-th]].
  
  
  
  

\bibitem{Alves:2014ksa} A.~Alves, A.~G.~Dias, R.~da Silva,
  Physica {\bf 420} (2015) 1 [{arXiv:1408.0827 [hep-ph]}].
 

\bibitem{Correa:2015vka} 
  R.~A.~C.~Correa and R.~da Rocha,
  Eur.\ Phys.\ J.\ C {\bf 75}, no. 11, 522 (2015)
  doi:10.1140/epjc/s10052-015-3735-8
  [arXiv:1502.02283 [hep-th]].
  
  
\bibitem{Sowinski:2017hdw} D.~Sowinski and M.~Gleiser,
  J.\ Stat.\ Phys.\  {\bf 167} (2017) no.5,  1221
  [{arXiv:1606.09641 [cond-mat.stat-mech]}].
 


\bibitem{Casadio:2016aum} R.~Casadio and R.~da Rocha,
   Phys. Lett. B {\bf 763} (2016) 434 [{arXiv:1610.01572 [hep-th]}].
   
   
 
\bibitem{Correa:2016pgr} R.~A.~C.~Correa, D.~M.~Dantas, C.~A.~S.~Almeida and R.~da Rocha,
  Phys.\ Lett.\ B {\bf 755} (2016) 358  
  [{arXiv:1601.00076 [hep-th]}].

  
  
\bibitem{Braga:2016wzx} 
  N.~R.~F.~Braga and R.~da Rocha,
  Phys.\ Lett.\ B {\bf 767}, 386 (2017)
  [arXiv:1612.03289 [hep-th]].

\bibitem{Karapetyan:2016fai} 
  G.~Karapetyan,
  EPL {\bf 117}, no. 1, 18001 (2017)
  [arXiv:1612.09564 [hep-ph]].
  

\bibitem{Karapetyan:2017edu} G.~Karapetyan,
  EPL {\bf 118} (2017) 38001 
  [{arXiv:1705.10617 [hep-ph]}].
 
 \bibitem{Alves:2017ljt} A.~Alves, A.~G.~Dias,  R.~da Silva,
 Braz. J. Phys. {\bf 47} (2017) 426 [{arXiv:1703.02061 [hep-ph]}].

   
\bibitem{Karapetyan:2018oye} G.~Karapetyan,
  Phys.\ Lett.\ B {\bf 781} (2018) 201 [{arXiv:1802.09105 [nucl-th]}].
  
   
\bibitem{Karapetyan:2018yhm}
  G.~Karapetyan,
  Phys.\ Lett.\ B {\bf 786} (2018) 418 [arXiv:1807.04540 [nucl-th]].
 
 
\bibitem{Gleiser:2018kbq} M.~Gleiser, M.~Stephens and D.~Sowinski,
  Phys.\ Rev.\ D {\bf 97}  (2018) 096007 
  [{arXiv:1803.08550 [hep-th]}].

\bibitem{Colangelo:2018mrt} 
  P.~Colangelo and F.~Loparco,
  Phys.\ Lett.\ B {\bf 788}, 500 (2019)
  doi:10.1016/j.physletb.2018.11.053
  [arXiv:1811.05272 [hep-ph]].


\bibitem{Lee:2018zmp}
  C.~O.~Lee,
  Phys.\ Lett.\ B {\bf 790} (2019) 197
  [arXiv:1812.00343 [gr-qc]].

\bibitem{Bazeia:2018uyg}
  D.~Bazeia, D.~C.~Moreira and E.~I.~B.~Rodrigues,
  J. Magn. Magn. Mater. {\bf 475} (2019) 734. 


\bibitem{Ma:2018wtw} 
  C.~W.~Ma and Y.~G.~Ma,
  Prog.\ Part.\ Nucl.\ Phys.\  {\bf 99}, 120 (2018)
  [arXiv:1801.02192 [nucl-th]].

 
\bibitem{Zhao:2019xle} 
  Q.~Zhao, B.~Z.~Mi and Y.~Li,
  Int.\ J.\ Mod.\ Phys.\ B {\bf 33}, no. 12, 1950119 (2019).
 
 
\bibitem{Ferreira:2019nkz} 
  L.~F.~Ferreira and R.~da Rocha,
  arXiv:1907.11809 [hep-th].

\bibitem{Bernardini:2019stn} 
  A.~E.~Bernardini and R.~da Rocha,
  Phys.\ Lett.\ B {\bf 796}, 107 (2019)
  doi:10.1016/j.physletb.2019.07.028
  [arXiv:1908.04095 [gr-qc]].
 
\bibitem{Karapetyan:2019ran} 
  G.~Karapetyan,
  EPL {\bf 129}, no. 1, 18002 (2020)
  doi:10.1209/0295-5075/129/18002
  [arXiv:1912.10071 [hep-ph]].
   


\bibitem{Bass:1998vz} 
  S.~A.~Bass, M.~Gyulassy, H.~Stoecker and W.~Greiner,
  J.\ Phys.\ G {\bf 25}, R1 (1999)
  doi:10.1088/0954-3899/25/3/013
  [hep-ph/9810281].
  
\bibitem{Scherer:1999qq} 
  S.~Scherer {\it et al.},
  Prog.\ Part.\ Nucl.\ Phys.\  {\bf 42}, 279 (1999).
  doi:10.1016/S0146-6410(99)00083-6
 
\bibitem{Shuryak:2008eq} 
  E.~Shuryak,
  Prog.\ Part.\ Nucl.\ Phys.\  {\bf 62}, 48 (2009)
  doi:10.1016/j.ppnp.2008.09.001
  [arXiv:0807.3033 [hep-ph]].
 
 
\bibitem{CasalderreySolana:2011us} 
  J.~Casalderrey-Solana, H.~Liu, D.~Mateos, K.~Rajagopal and U.~A.~Wiedemann,
  book:Gauge/String Duality, Hot QCD and Heavy Ion Collisions. Cambridge, UK: Cambridge University Press, 2014
  doi:10.1017/CBO9781139136747
  [arXiv:1101.0618 [hep-th]].
 
\bibitem{Braga:2015jca} 
  N.~R.~F.~Braga, M.~A.~Martin Contreras and S.~Diles,
  Phys.\ Lett.\ B {\bf 763}, 203 (2016)
  [arXiv:1507.04708 [hep-th]].
 
 \bibitem{Braga:2016wkm} 
  N.~R.~F.~Braga, M.~A.~Martin Contreras and S.~Diles,
  Eur.\ Phys.\ J.\ C {\bf 76}, no. 11, 598 (2016)
  [arXiv:1604.08296 [hep-ph]].
      
 
 
 
\bibitem{Braga:2017oqw} 
  N.~R.~F.~Braga and L.~F.~Ferreira,
  Phys.\ Lett.\ B {\bf 773}, 313 (2017)
  [arXiv:1704.05038 [hep-ph]].
 
 
\bibitem{Braga:2017bml} 
  N.~R.~F.~Braga, L.~F.~Ferreira and A.~Vega,
  Phys.\ Lett.\ B {\bf 774}, 476 (2017)
  [arXiv:1709.05326 [hep-ph]]. 

 
\bibitem{Braga:2018zlu} 
  N.~R.~F.~Braga and L.~F.~Ferreira,
  Phys.\ Lett.\ B {\bf 783}, 186 (2018)
  [arXiv:1802.02084 [hep-ph]].
 
\bibitem{Braga:2018hjt} 
  N.~R.~F.~Braga and L.~F.~Ferreira,
  JHEP {\bf 1901}, 082 (2019)
  [arXiv:1810.11872 [hep-ph]].
 
 
\bibitem{Braga:2019yeh} 
  N.~R.~F.~Braga and L.~F.~Ferreira,
  Phys.\ Lett.\ B {\bf 795}, 462 (2019)
  [arXiv:1905.11309 [hep-ph]].
 
  
\bibitem{Braga:2019xwl} 
  N.~R.~F.~Braga and R.~Da Mata,
  arXiv:1910.13498 [hep-ph].


\bibitem{shannon}
 C. E. Shannon, The Bell System Technical Journal, {\bf 27}, 379 (1948). 
 
 
 \bibitem{Birula}   Białynicki-Birula, I. and Mycielski,  
 Commun.Math. Phys. 44, 129–132 (1975) 
 https://doi.org/10.1007/BF01608825. 
 
 \bibitem{Braga:2019jqg} 
  N.~R.~F.~Braga,
  Phys.\ Lett.\ B {\bf 797}, 134919 (2019)
  doi:10.1016/j.physletb.2019.134919
  [arXiv:1907.05756 [hep-th]].
 
 
\bibitem{Colangelo:2010pe} 
  P.~Colangelo, F.~Giannuzzi and S.~Nicotri,
  Phys.\ Rev.\ D {\bf 83}, 035015 (2011)
  doi:10.1103/PhysRevD.83.035015
  [arXiv:1008.3116 [hep-ph]].
  
\bibitem{Colangelo:2011sr} 
  P.~Colangelo, F.~Giannuzzi, S.~Nicotri and V.~Tangorra,
  Eur.\ Phys.\ J.\ C {\bf 72}, 2096 (2012)
  doi:10.1140/epjc/s10052-012-2096-9
  [arXiv:1112.4402 [hep-ph]].
  
\bibitem{Colangelo:2012jy} 
  P.~Colangelo, F.~Giannuzzi and S.~Nicotri,
  JHEP {\bf 1205}, 076 (2012)
  doi:10.1007/JHEP05(2012)076
  [arXiv:1201.1564 [hep-ph]].
 
 
    
  
\bibitem{Branz:2010ub} 
  T.~Branz, T.~Gutsche, V.~E.~Lyubovitskij, I.~Schmidt and A.~Vega,
  Phys.\ Rev.\ D {\bf 82}, 074022 (2010)
  [arXiv:1008.0268 [hep-ph]].
  
\bibitem{Gutsche:2012ez} 
  T.~Gutsche, V.~E.~Lyubovitskij, I.~Schmidt and A.~Vega,
  Phys.\ Rev.\ D {\bf 87}, no. 5, 056001 (2013)
  [arXiv:1212.5196 [hep-ph]].
  
  \bibitem{Fadafan:2011gm} 
  K.~B.~Fadafan,
  Eur.\ Phys.\ J.\ C {\bf 71}, 1799 (2011)
  doi:10.1140/epjc/s10052-011-1799-7
  [arXiv:1102.2289 [hep-th]].
  
  \bibitem{Fadafan:2012qy} 
  K.~B.~Fadafan and E.~Azimfard,
  Nucl.\ Phys.\ B {\bf 863}, 347 (2012)
  doi:10.1016/j.nuclphysb.2012.05.022
  [arXiv:1203.3942 [hep-th]].
  
  \bibitem{Fadafan:2013coa} 
  K.~B.~Fadafan and S.~K.~Tabatabaei,
  Eur.\ Phys.\ J.\ C {\bf 74}, 2842 (2014)
  doi:10.1140/epjc/s10052-014-2842-2
  [arXiv:1308.3971 [hep-th]].
  
  \bibitem{Afonin:2013npa} 
  S.~S.~Afonin and I.~V.~Pusenkov,
  Phys.\ Lett.\ B {\bf 726}, 283 (2013)
  [arXiv:1306.3948 [hep-ph]].
  
  
\bibitem{Mamani:2013ssa} 
  L.~A.~H.~Mamani, A.~S.~Miranda, H.~Boschi-Filho and N.~R.~F.~Braga,
  JHEP {\bf 1403}, 058 (2014)
  doi:10.1007/JHEP03(2014)058
  [arXiv:1312.3815 [hep-th]].
  
  
  
  
  
  \bibitem{Hashimoto:2014jua} 
  K.~Hashimoto, N.~Ogawa and Y.~Yamaguchi,
  JHEP {\bf 1506}, 040 (2015)
  [arXiv:1412.5590 [hep-th]].
  
  
\bibitem{Liu:2016iqo} 
  Y.~Liu and I.~Zahed,
  Phys.\ Rev.\ D {\bf 95}, no. 5, 056022 (2017)
  doi:10.1103/PhysRevD.95.056022
  [arXiv:1611.03757 [hep-ph]].
  
  
\bibitem{Liu:2016urz} 
  Y.~Liu and I.~Zahed,
  Phys.\ Lett.\ B {\bf 769}, 314 (2017)
  doi:10.1016/j.physletb.2017.04.007
  [arXiv:1611.04400 [hep-ph]].
  
\bibitem{Ballon-Bayona:2017bwk} 
  A.~Ballon-Bayona, G.~Krein and C.~Miller,
  Phys.\ Rev.\ D {\bf 96}, no. 1, 014017 (2017)
  doi:10.1103/PhysRevD.96.014017
  [arXiv:1702.08417 [hep-ph]].
  
   
  
\bibitem{Dudal:2018rki} 
  D.~Dudal and T.~G.~Mertens,
  Phys.\ Rev.\ D {\bf 97}, no. 5, 054035 (2018)
  doi:10.1103/PhysRevD.97.054035
  [arXiv:1802.02805 [hep-th]].
   
    
   
\bibitem{Gutsche:2019blp} 
  T.~Gutsche, V.~E.~Lyubovitskij, I.~Schmidt and A.~Y.~Trifonov,
  Phys.\ Rev.\ D {\bf 99}, no. 5, 054030 (2019)
  doi:10.1103/PhysRevD.99.054030
  [arXiv:1902.01312 [hep-ph]].
   
\bibitem{Bohra:2019ebj} 
  H.~Bohra, D.~Dudal, A.~Hajilou and S.~Mahapatra,
  arXiv:1907.01852 [hep-th].
 
\bibitem{Zhang:2019qhm}
  Z.~q.~Zhang and X.~Zhu,
  Phys.\ Lett.\ B {\bf 793} (2019) 200.
  doi:10.1016/j.physletb.2019.04.057
   
\bibitem{MartinContreras:2019kah} 
  M.~A.~Martin Contreras and A.~Vega,
  Phys.\ Rev.\ D {\bf 101}, no. 4, 046009 (2020)
  doi:10.1103/PhysRevD.101.046009
  [arXiv:1910.10922 [hep-th]].
    
\end{thebibliography}
 \end{document}